\documentclass[sigconf]{acmart}
\usepackage{comment}

\usepackage[show]{imports/chato-notes}
\usepackage[utf8]{inputenc}

\renewcommand\footnotetextcopyrightpermission[1]{} 
\usepackage{booktabs} 

\setcopyright{none}

\usepackage{MnSymbol}
\usepackage{amsfonts}
\usepackage{proof}
\usepackage{algorithm}
\usepackage{algpseudocode}
\usepackage{listings}
\usepackage{subfig}
\usepackage{fancybox}
\usepackage{xspace}
\usepackage{cleveref}
\usepackage{tabularx}
\usepackage{booktabs}
\usepackage{multirow}
\usepackage{makecell}
\usepackage{caption}
\usepackage{array}
\usepackage{icomma}
\usepackage{graphicx}
\usepackage[figuresright]{rotating}
\usepackage{afterpage}
\usepackage{printlen}

\DeclareGraphicsExtensions{.pdf,.png,.jpg}

\newcommand{\eat}[1]{}

\algrenewcommand\algorithmicindent{1em}%

\newcommand*{\cyclerank}{{\slshape{CycleRank}}\xspace}
\newcommand*{\ppagerank}{{\slshape{PPageRank}}\xspace}
\newcommand*{\categories}{{\slshape{Categories}}\xspace}

\newcommand*{\wiki}[1]{\emph{``#1''}\xspace}

\usepackage[table,xcdraw]{xcolor}

\begin{document}
\title[A general method for estimating the prevalence of...]{A general method for estimating the prevalence of Influenza-Like-Symptoms with Wikipedia data}

\author{Giovanni De Toni}
\affiliation{%
  \institution{DISI, University of Trento}
}
\email{giovanni.detoni@unitn.it}

\author{Cristian Consonni}
\affiliation{%
  \institution{Eurecat - Centre Tecnòlogic of Catalunya}
}
\email{cristian.consonni@eurecat.org}

\author{Alberto Montresor}
\affiliation{%
  \institution{DISI, University of Trento}
}
\email{alberto.montresor@unitn.it}

\renewcommand{\shortauthors}{G. De Toni et al.}

\maketitle

\section*{Abstract}
Influenza is an acute respiratory seasonal disease that affects millions of people worldwide and causes thousands of deaths in Europe alone. Being able to estimate in a fast and reliable way the impact of an illness on a given country is essential to plan and organize effective countermeasures, which is now possible by leveraging unconventional data sources like web searches and visits. In this study, we show the feasibility of exploiting information about Wikipedia’s page views of a selected group of articles and machine learning models to obtain accurate estimates of influenza-like illnesses incidence in four European countries: Italy, Germany, Belgium, and the Netherlands. We propose a novel language-agnostic method, based on two algorithms, \textit{Personalized PageRank} and \cyclerank, to automatically select the most relevant Wikipedia pages to be monitored without the need for expert supervision. We then show how our model is able to reach state-of-the-art results by comparing it with previous solutions.


\section{Introduction}\label{section:introduction}

Influenza is a widespread acute respiratory infection that occurs with seasonal frequency, mainly during the autumn-winter period. The European Centre for Disease and Control (ECDC) estimates that  influenza causes up to 50 millions cases and up to $70,000$ deaths annually in Europe~\cite{ECDC2018,thompson2003mortality}. Globally, the number of deaths caused by this infection ranges from $290,000$ to $650,000$~\cite{WHO2019}. The most affected categories are children in the pediatric age range and seniors. Severe cases are recorded in people above 65 years old and with previous health conditions, for example, respiratory or cardiovascular illnesses~\cite{thompson2009estimates,nair2011global}.

The ECDC monitors the influenza virus activities and provides weekly bulletins that aggregate the open data coming from all the European countries. The local data are gathered from the national networks of volunteer physicians~\cite{ECDC2017}. The Centre supplies only data about the current situation, which means that they do not provide any valuable information about which will be the future impact of the influenza illness in a European country, namely the number of future infected people. Moreover, ECDC data are available with a delay ranging between one and two weeks, which prevents taking preventive measures to mitigate the impact of influenza.
In recent years, thanks to machine learning techniques, new methods have been developed to estimate the activity level of diseases by harnessing the power of the internet and big data. Researchers tried to use unconventional sources of data to make predictions, for example, tweets from Twitter~\cite{signorini2011use,santos2014analysing,kim2013use}, Google's search keywords~\cite{yang2015accurate} or social media~\cite{woo2016estimating}.

Wikipedia\footnote{\url{https://www.wikipedia.org}} is a multilingual encyclopedia, written collaboratively by volunteers online. As of this writing, the English version alone has more than 6M articles and 50M pages and it is edited on average by more than 65k active users every month~\cite{wikipedia2016statistics}.

Research has shown that Wikipedia is a first-stop source for information of all kinds, including information about
science~\cite{yasseri2014most, spoerri2007popular}. This is of particular significance in fields such as medicine,
where it has been shown that Wikipedia is a prominent result for over $70\%$ of search results for health-related keywords~\cite{laurent2009seeking}. Wikipedia is also the most viewed medical resources globally~\cite{heilman2015wikipedia}, and it is used as an information source by $50\%$ to $70\%$ of practicing physicians~\cite{heilman2011wikipedia}.

In this study, we investigate how to exploit Wikipedia as a source of data for doing rapid now-casting of the influenza activity levels in several European countries. McIver and Brownstein~\cite{mciver2014wikipedia} developed a method based on Wikipedia's pageviews, which behaves quite well when predicting influenza levels in the United States provided by the Centers for Disease Control and Prevention (CDC). They use pageviews of a predefined set of pages from the English language edition of Wikipedia chosen by a panel of experts. Generous et al.~\cite{generous2014global} also developed linear machine learning models and demonstrated the ability to forecast incidence levels for 14 location-disease combinations.

We extend upon the state of the art by proposing a novel method that can be applied automatically to other languages and countries. We focused our attention on predicting influenza levels in Italy, Germany, Belgium and the Netherlands. In extending this work, we have developed a method that is more flexible with respect of the characteristics of the edition of Wikipedia and that can be applied to any language thus enabling us to build and deploy machine learning models for \textit{Influenza-like Illnesses (ILI)} estimation without the need of expert supervision. 

The paper is structured as follows: in Section \ref{sec:mm} we review the relevant datasets that we have used in our study. In Section~\ref{sec:mm:methods} and ~\ref{sec:ml_models} we discuss our models and how we selected the relevant features to use for predicting influenza activity levels from Wikipedia data. We present the results of our study in Section~\ref{sec:results} and \ref{sec:results-features}, and we discuss them in Section~\ref{sec:discussion}. Finally, we conclude the paper in Section~\ref{sec:conclusions}.

\section{Data sources}\label{sec:mm}

\paragraph*{Wikipedia page views}\label{sss:wikipedia_pageviews}
Data about Wikipedia page views are freely available to download. For our study, we have used two datasets: the \texttt{pagecounts} dataset\footnote{Pagecounts-raw on \url{wikitech.wikimedia.org}, \url{https://w.wiki/gnq}, visited on 2020-10-14} (December 2007-May 2016, for a total of 461 weeks) and the \texttt{pageviews} dataset (October 2016 to April 2019, for a total of 134 weeks).
Both of them count the number of page hits to Wikipedia articles. The \texttt{pagecounts} dataset contains non-filtered page views, including automatically bot-generated ones; moreover, it does not contain data of page hits from mobile devices. 
The \texttt{pageviews} dataset has been recently developed, superseding \texttt{pagecounts} in October 2016, and includes only human traffic from both desktop and mobile devices. By considering both these datasets, we can examine a longer time range while estimating the impact of different models on our results.

From these datasets, we extracted the page views of three different versions of Wikipedia: Italian, German, and Dutch. We selected these languages because the majority of page views for each of them come almost exclusively from one single area.\footnote{Report: Breakdown of page views by language on \url{stats.wikimedia.org}, \url{https://w.wiki/gnr} visited on 2020-10-14}
For instance, just $41.5\%$ of the page views of the English Wikipedia come from US and UK, while all the rest comes from all over the world. On the contrary, $87.0\%$ of the page views of the Italian version of Wikipedia come from Italy, $77.0\%$ of the page views of the German version of Wikipedia come from Germany, and almost $89\%$ of the Dutch Wikipedia's page views come from the Netherlands and Belgium ($20.3\%$ coming from Belgium and $69.1\%$ coming from the Netherlands alone). This enables us to build datasets which contain less noise and which are only affected by the internet behaviour of the population of the country or area we are examining.

Each of row is composed of four columns, namely: the project's language, the page title, the number of requests and the size in byte of the page retrieved. Data are available with hourly granularity; to use them in our project, we filtered them to select only specific entries and then we computed an aggregated weekly value of the page views for each of the entries selected.

We generated weekly aggregates for page views for the period ranging from the 50th week of 2007 to the 15th week of 2019, for a total of 591 weeks. \texttt{pageviews} data were used to cover the period from September 2016 to April 2019 period, while for the previous period  \texttt{pagecounts} data were used.

We processed weekly page views by standardizing them, i. e. removing the mean and then dividing by the standard deviation. In case of missing data, we used a padding technique to propagate the last valid value to fill the empty entries. Since Wikipedia articles are created collaboratively by the community of editors, their creation date may vary: we assigned the value of zero to the page view count to the period when they did not exist.
We included the week number as a feature of our model. In order to do so, we one-hot encoded the week value into binary vectors $w_i \in \{0, 1\}^{52}$ where the bit corresponding to the given week is set to $1$. 

We then built two different datasets named \texttt{PV} and \texttt{PC+PV}. \texttt{PV}  contains only the \texttt{pageviews} data, thus covering only four influenza seasons (from 2015-2016 to 2018-2019). \texttt{PC+PV} was obtained by merging \texttt{pagecounts} and \texttt{pageviews} data, thus covering all the influenza seasons (from 2007-2008 to 2018-2019).

\paragraph*{Influenza activity levels}\label{sec:mm:data}
Data about the incidence of influenza in the population are available through different online official websites and repositories. We collected influenza data for these countries: Italy, Germany, Austria and Netherlands. In the following we describe briefly the collection procedure we followed for each country. 

Italian data were downloaded from the InfluNet system\footnote{InfluNet on \url{epicentro.iss.it},  \url{https://www.epicentro.iss.it/influenza/influnet}, visited on 2020-10-15}, which is the Italian flu surveillance program managed by the Istituto Superiore di Sanità; German data were obtained from the Robert Koch Institute\footnote{Robert Koch Institute, on \url{www.rki.de}, \url{https://www.rki.de/DE/Home/homepage\_node.html}, visited on 2020-10-20}; Dutch and Belgian data were take from the GISRS (Global Influenza Surveillance and Response System) online tool which is available from the WHO website\footnote{WHO Influenza Surveillance on \url{www.who.it/influenza/en}, \url{https://www.who.int/influenza/en/}, visited on 2020-10-20}, since at the best of our knowledge, the two countries do not offer open data from national health institute websites. 

For Germany and Italy, we cover 12 influenza seasons (from 2007-2008 to 2018-2019); for the Netherlands and Belgium, we have data that cover 9 influenza seasons (from 2010-2011 to 2018-2019). Each of the datasets records the influenza incidence aggregated weekly over 100\,000 patients. For each influenza seasons, each of the dataset entries represents the influenza level on a specific week.

For each influenza seasons, data cover 26 weeks, from the 42nd week of the previous year to the 15th week of the following year; for instance, the 2009-2010 influenza season comprises data ranging from the 42nd week of 2009 to the 15th week of 2010. The complete dataset has the following size: 312 weeks of data for Germany and Italy, 234 for the Netherlands and Belgium.

\begin{table*}[t]
  \caption{Selected Wikipedia categories. We first selected the Italian categories and then we chose the corresponding German and Dutch translation. The English categories are reported for translation purposes.}
  \label{tab:wikipedia_categories}
  \begin{tabular}{@{}llll@{}}
  \toprule
    \textbf{English}
  & \textbf{Italian} 
  & \textbf{German}
  & \textbf{Dutch}
  \\ \midrule
    Viral diseases
  & Malattie infettive virali
  & Virusziekte
  & Virale Infektionskrankheit
  \\
    Infectious diseases
  & Malattie infettive       
  & Infectieziekte
  & Infektionskrankheit
  \\
    Epidemics
  & Epidemie                 
  & Epidemie
  & Epidemie
  \\
    Viruses
  & Virus                    
  & Virus
  & Viren, Viroide und Prionen
  \\
    Vaccines
  & Vaccini                  
  & Vaccin
  & Impfstoff
  \\
    Medical signs
  & Segni clinici
  & Symptoom
  & Krankheitssymptom
  \\ \bottomrule
\end{tabular}
\end{table*}

\section{Feature Selection}\label{sec:mm:methods}

The selection of the pages to monitor was based on the following assumption: during the influenza seasons, we should see an increment in the page view number of some specific Wikipedia's pages, which should be related to flu topics (e.g., \wiki{Influenza}, \wiki{Headache}, \wiki{Flu vaccine}, etc.). That occurs because most people tend to search online the symptoms of their disease when they are sick, trying to get more information about their illness.

We use three different methodologies in order to avoid to make unnecessary assumptions about which pages to choose. 

\subsection{Exploiting Wikipedia's Categories}

The first method consists of determining a series of Wikipedia's categories related to the medical sector, from which to extract the single entries. We started by choosing the categories from the Italian version of Wikipedia. After doing that, we replicated the procedure for all the other languages (German, Dutch) by selecting the corresponding translation of the given category in the Italian Wikipedia. We ended up having a dedicated set of Wikipedia's pages for each of the different Wikipedia's versions. In the end, we produced three different lists of Wikipedia's pages which were used to filter the Wikipedia's page view dumps. 

The categories selected by this method can be seen in Table~\ref{tab:wikipedia_categories}.

\subsection{Graph-based methods}

The strategies that we propose in the following section can choose automatically the set of relevant pages to monitor. In this way, we could keep our page list always up-to-date with minimum effort, yet retaining its effectiveness. 

Wikipedia can be represented as a graph in which each article is a node and links between them are directed edges. Wikipedia editors write articles and insert link to other pages which are contextually relevant; links are also used by readers to navigate between pages. When choosing which pages are relevant we should consider the entire graph structure; for example, we may consider the pages linked by the \wiki{Influenza} article to be relevant, but also the ones that link to it. Previous work has showed that Wikipedia links provide a way to explore the context of an article~\cite{west2015} and we can exploit this knowledge to extract the best pages to monitor. Ideally, we would like to be able to give to each Wikipedia page a score that measures its relative importance to a given ``source'' page, in this case the \wiki{Influenza} page for each language.

There are several algorithms that can be used to given scores to graph nodes based on specific metrics. We adopt here two approaches: the well-known Personalized PageRank~\cite{page1999pagerank} and a more recent proposal called CycleRank~\cite{consonni2020}.

\subsubsection*{Personalized PageRank} 

The Personalized PageRank algorithm~\cite{page1999pagerank} (\ppagerank for short) ranks web pages in order of importance with respect to a set of \textit{source pages}. \ppagerank models the relevance of nodes around the selected nodes as the probability of reaching each of them, when following random walks starting from one of the sources nodes. 

This algorithm has been applied successfully to a broad range of graph structures in order to find relevant items~\cite{doi:10.1137/140976649, spark2019} and we include it as a baseline in our analysis. Its performance, however, is hindered by pages with high in-degree that function as hubs and obtain high scores regardless of the starting point.

\subsubsection*{CycleRank}\label{sec:methods:cyclerank}

The \cyclerank algorithm~\cite{consonni2020} is a novel approach based on cycles, i.e. circular walks that starts and end in a given node, exploiting both incoming and outgoing links to reduce the importance of in-degree hubs. 
The idea behind this method is that by using circular walks, it is possible to identify pages that are more pertaining to the context of the topic of interest. Cycles guarantee that we rank higher pages that are, at least indirectly, both linked from and linking to the reference article, thus avoiding the pitfalls of \ppagerank where random walks may easily lead far away from the original topic, into pages that are not relevant at all.

\begin{table*}[ht!]
    \centering
    \caption{Percentages of features shared by each feature set given the method used to extract them. Since the various datasets have different size, the percentages are not symmetrical. Each row shows the result for one of the methods. For instance, the \cyclerank row shows the fraction of features it shares with the other feature sets.}
    \begin{tabular}{@{}c|ccc|ccc|ccc|@{}}
\cline{2-10}
\multicolumn{1}{l|}{} & \multicolumn{3}{c|}{\textbf{CycleRank}} & \multicolumn{3}{c|}{\textbf{PPageRank}} & \multicolumn{3}{c|}{\textbf{Categories}} \\ \cline{2-10} 
\multicolumn{1}{l|}{} & \textit{Italian} & \textit{German} & \textit{Dutch} & \textit{Italian} & \textit{German} & \textit{Dutch} & \textit{Italian} & \textit{German} & \textit{Dutch} \\ \cline{2-10} 
\textbf{CycleRank} & 100\% & 100\% & 100\% & 53,75\% & 49,36\% & 83,72\% & 18,75\% & 10,97\% & 41,86\% \\
\textbf{PPageRank} & 35,10\% & 42,23\% & 26,27\% & 100\% & 100\% & 100\% & 10,20\% & 8,30\% & 15,32\% \\
\textbf{Categories} & 6,38\% & 6,82\% & 9,42\% & 5,31\% & 6,04\% & 10,99\% & 100\% & 100\% & 100\% \\ \cline{2-10}
\end{tabular}
    \label{tab:percentage-feature-shared}
\end{table*}

The \cyclerank score can be seen as the time spent on a given node when following random loops from the reference node, assuming a fixed overall time for each loop, equally split among all the nodes encountered. The \cyclerank algorithm is similar to \ppagerank, because it can be explained as the probability of ending up in a node when following a random path that starts and ends in the source nodes. 

\subsection{Discussion}

We then analyze the set of features extracted by using these different methodologies. For the Dutch, German and Italian Wikipedia versions, respectively, the \categories feature sets contain 382, 381 and 470 Wikipedia's pages; the \ppagerank feature sets contain 274, 277 and 245 pages; and the \cyclerank feature sets contain 86, 237 and 160 pages.

The feature sets built from the categories contain more pages than the other two, this is expected since we do not impose any constraint on them. The \cyclerank and \ppagerank feature sets are smaller since we enforce that all those pages must be connected to the \wiki{Influenza} page; \cyclerank sets are the smallest ones because they require connections in both directions. In general, the \categories feature set encompass a broad spectrum of pages related to medical topics, which may or may not have any relation with the \wiki{Influenza} concept.

Table \ref{tab:percentage-feature-shared} shows how much the features sets overlaps across the three methods under consideration. \cyclerank and \ppagerank feature lists share many common pages. For instance, $42\%$ of the pages selected by \cyclerank are also selected by the \ppagerank, while \categories feature sets are more diverse, with many pages appearing only with this method. However, for the Dutch Wikipedia, \cyclerank shares $42\%$ of its features with the \categories Dutch feature set.

At a first glance, \textit{Categories} method outlined above is difficult and prone to errors. Choosing the correct pages to monitor is a time-consuming task which requires some expertise about the medical and epidemiological domain. Moreover, creating and maintaining this list of pages is a process which needs to be done manually and frequently. Wikipedia's structure can undergo rapid changes and each of its pages can be deleted, moved or renamed, thus forcing us to keep our list of features constantly updated . On the other hand, \cyclerank and \ppagerank do not require human intervention and they are surely faster than the manual \textit{Categories}.

\section{Models}\label{sec:ml_models}

We trained a simple linear model regularized with the least absolute shrinkage and selection (LASSO) method \cite{tibshirani1996regression}. The LASSO regularization diminishes the possible over-fitting and automatically reduces the number of features used by the models. The models were trained by stochastic gradient descent (SGD) and by trying to minimize the Mean Squared Error (MSE). During training, we used cross-validation to estimate the perfect balance for the LASSO regularization term.
We tried to use the simplest technique possible, linear regression, because we assumed that the correlation between the influenza incidence and the page views of Wikipedia was linear. Moreover, we can use the linear model's weights to visualize the relative importance of each page towards the final prediction. 

We tested and evaluated three different linear models: \categories, \ppagerank and \cyclerank, each reflecting the set of features used. Each model was trained using the \texttt{PV} and \texttt{PC+PV} datasets mentioned above.

The models were trained on all the influenza seasons by leaving out the one we wanted to estimate. Basically, we trained a different model for each influenza season. We computed an estimate of the final performances as an average over all the trained models. Apart from the last year, every model has been evaluated on a given influenza season (e.g., 2015-2016), but the model has been trained by using also future influenza season data (e.g., 2016-2017, 2017-2018, etc.). We argue that this is acceptable since we assume the various influenza period to be conditionally independent. Despite many factors could make this assumption less precise, we claim that this experimental setting is still appropriate to give us an idea about the performance of the final complete models, which will be used to estimate ILI case on new data.  

The code\footnote{\url{https://github.com/fluTN/influenza}} and the dataset~\cite{detoni2020fluTN} used for the project are released as an open-source project to ensure correct peer-review and to assure reproducibility of our results.


\section{Results}\label{sec:results}

We compared the performance of the \cyclerank and \ppagerank models against the one trained with the user-defined categories (\categories). We trained the models to predict the last four influenza seasons (2015-2016, 2016-2017, 2017-2018 and 2018-2019). This analysis was done for all the chosen European countries: Italy (IT), Germany (DE), Belgium (BE) and the Netherlands (NL). We measured the Pearson Correlation Coefficient (PCC) to determine which of the models held the best performances by taking the mean of the various PCC obtained for each of the four seasons. 

We also evaluated the accuracy in estimating the week in which the influenza peak occurs (highest incidence value over the entire season). We considered a peak estimation to be acceptable if the model was able to estimate the peak with a deviation of $\pm 2$ weeks with respect to the ground truth. The models were trained by using the \texttt{PV} and \texttt{PC+PV} datasets.
Namely, we wanted to analyze the effects of providing additional data during training, even if it was recorded with a different methodology. 
In this work we also tried to evaluate the goodness of the most important features selected by the various models. We tried to understand which Wikipedia's pages are the most valuable when estimating the influenza incidence over a population. Moreover, we wanted also to assess the effectiveness of our automatic method with respect to the traditional, manual one (\categories). 

\subsection{Estimation Accuracy}

Table~\ref{tab:experiments-results} shows the complete results for each model and each country. All models present a high PCC with the influenza incidence data. Generally, \cyclerank and \ppagerank yield better results or provide similar performances to the \categories model. For instance, the latter shows lower performances on Germany and the Netherlands (both with the \texttt{PV} and \texttt{PC+PV} datasets). 
The \ppagerank model scores the highest correlation coefficient on 3 out of 4 of the examined countries (IT, DE, BE) with the best model for NL being \cyclerank. In general, \ppagerank and \cyclerank show a lower variance in PCC than the \categories model.

Examining the effect of using different datasets to train the various models, in 8 over 12 experiments, the \texttt{PC+PV} dataset improves the PCC of about $10\%$ on average. This effect is visible for \cyclerank and \categories, while  \ppagerank seems to not obtain the same gain when employing a larger page view dataset.  
In the remaining four experiments, we noticed that using a larger training dataset produced an average accuracy loss of around $5\%$.

\begin{figure*}[hbtp]
\centering

\subfloat{
\includegraphics[width=0.9\linewidth]{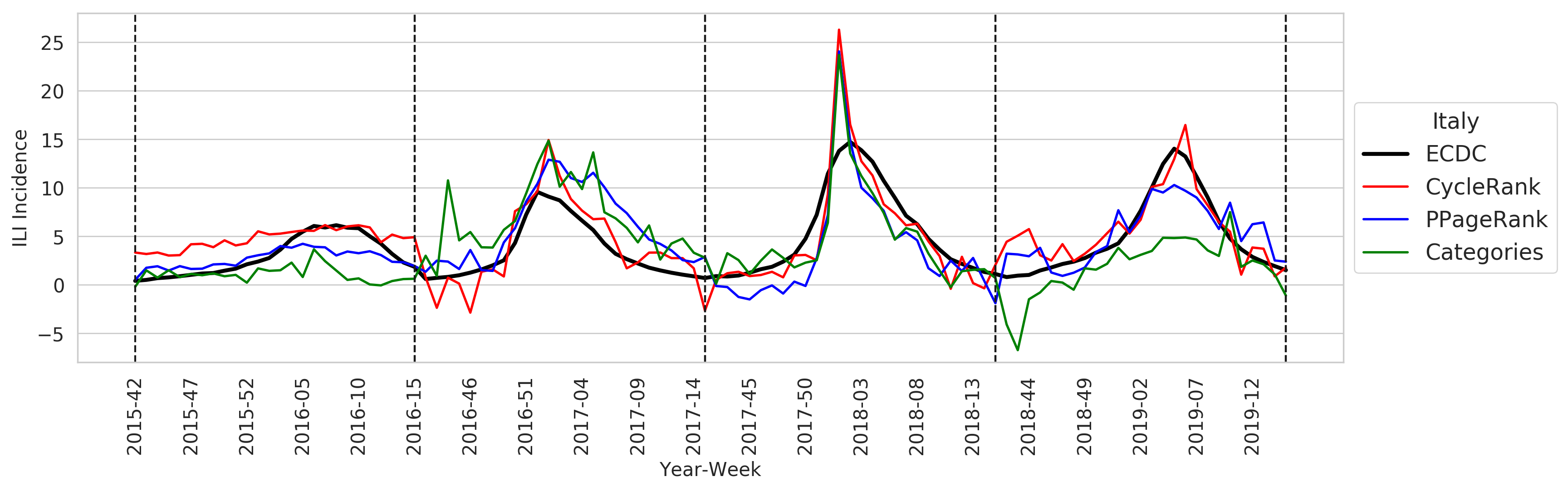}
}

\subfloat{
\includegraphics[width=0.9\linewidth]{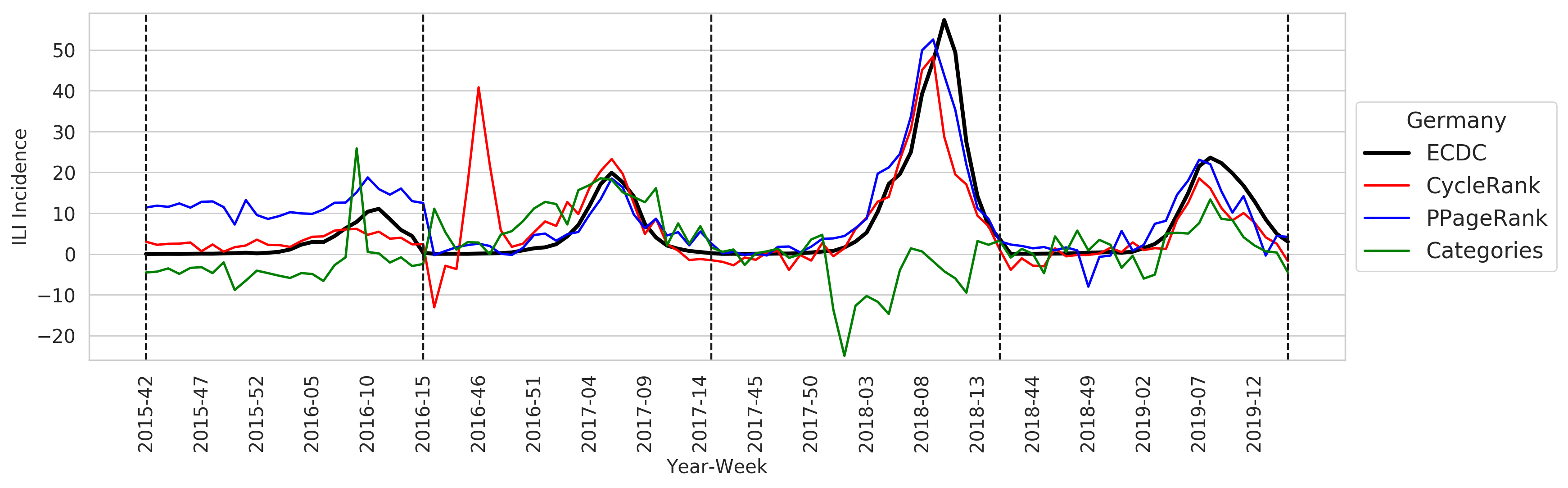}
}

\subfloat{
\includegraphics[width=0.9\linewidth]{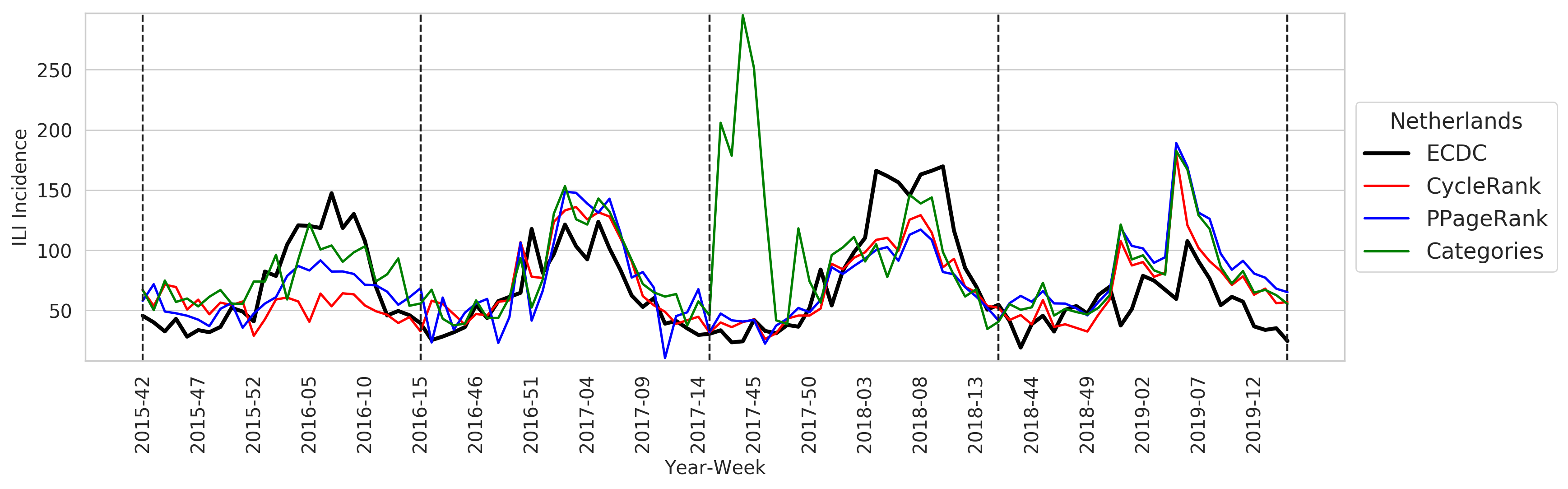}
}

\subfloat{
\includegraphics[width=0.9\linewidth]{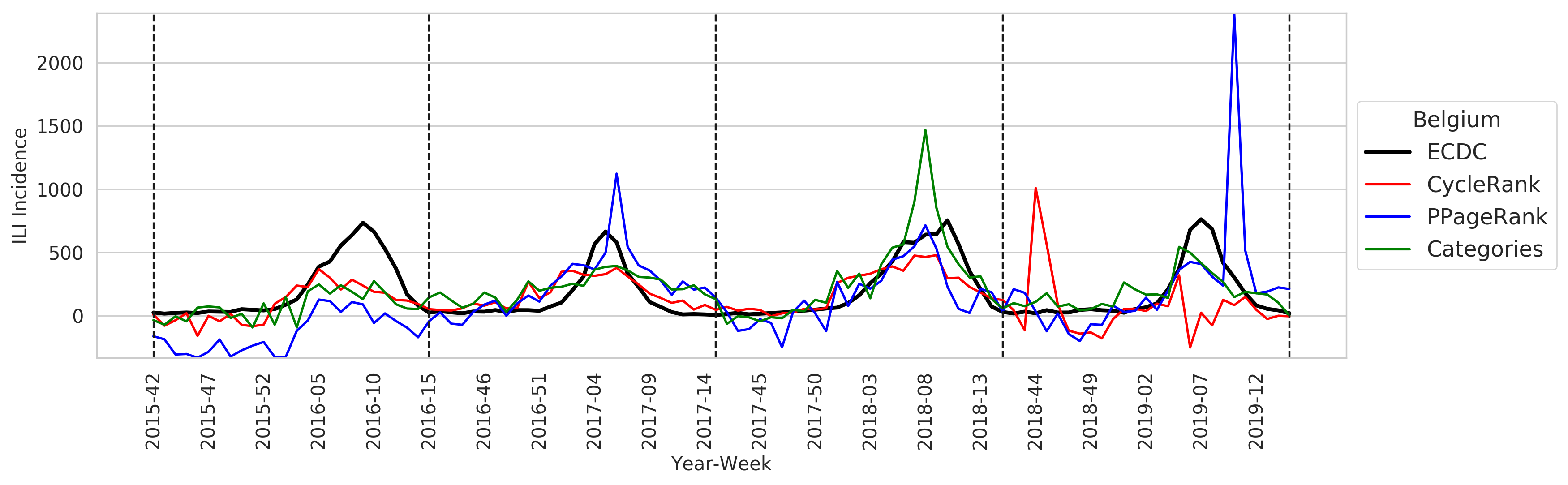}
}

\caption{\cyclerank, \ppagerank and \categories models predictions using the \texttt{PV} dataset on the Italian, German, Belgian and Dutch influenza incidence. The dashed lines delimits the various influenza seasons.}
\label{fig:model-prediction-pageviews}

\end{figure*}

\begin{figure*}[hbtp]
\centering

\subfloat{
\includegraphics[width=0.9\linewidth]{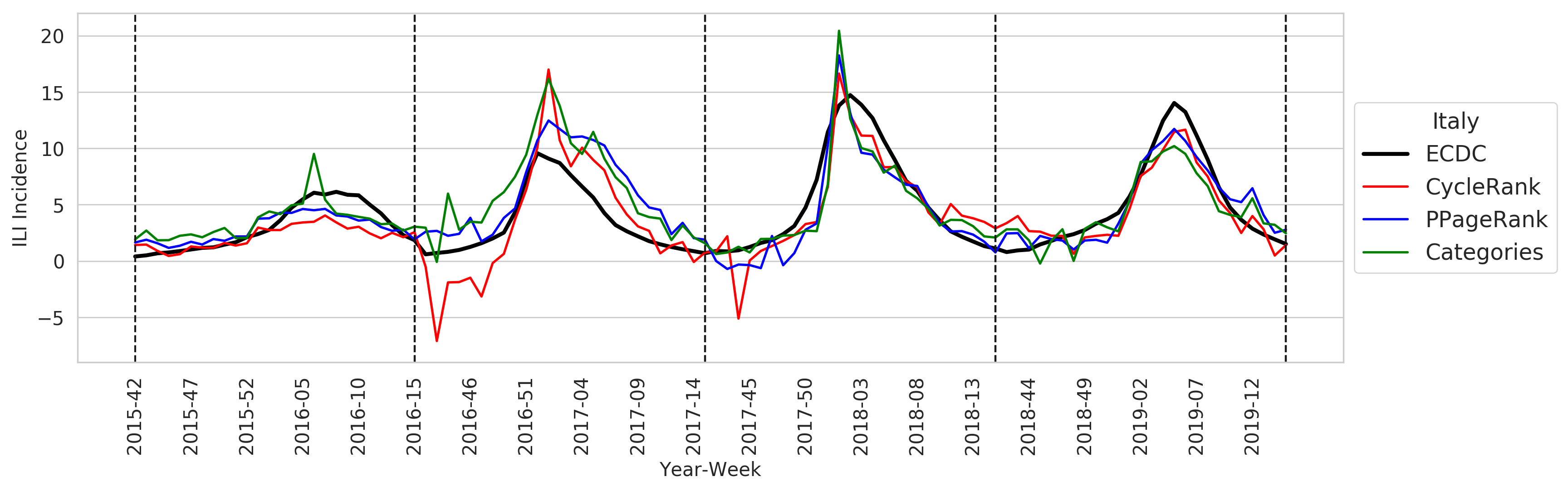}
}

\subfloat{
\includegraphics[width=0.9\linewidth]{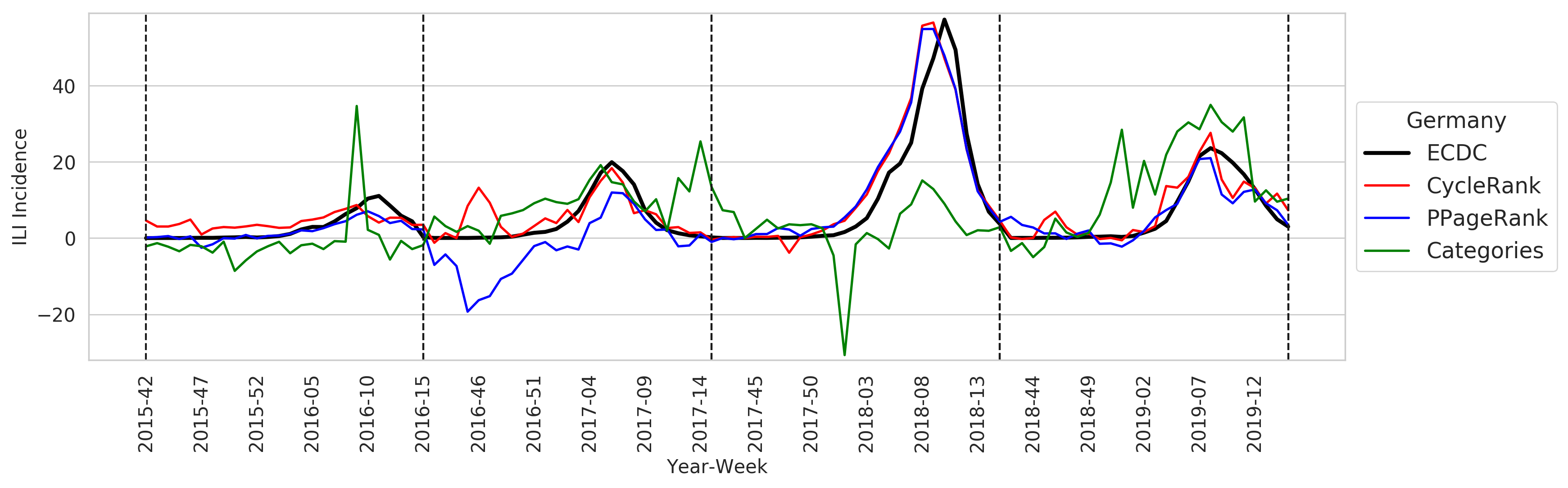}
}

\subfloat{
\includegraphics[width=0.9\linewidth]{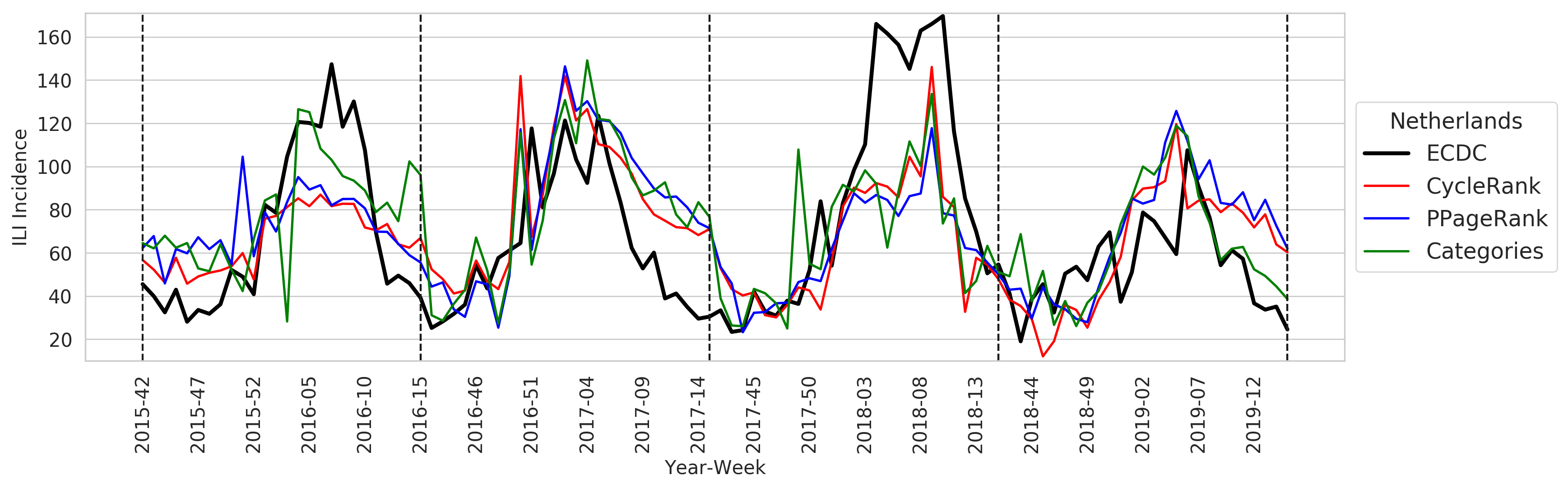}
}

\subfloat{
\includegraphics[width=0.9\linewidth]{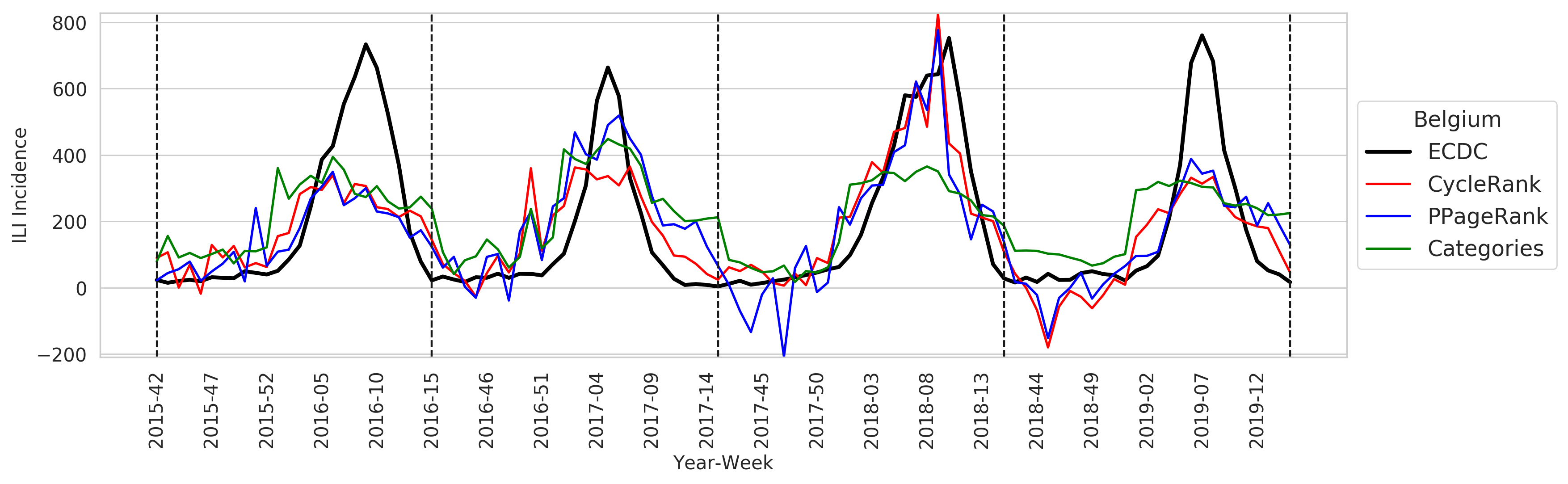}
}

\caption{\cyclerank, \ppagerank and \categories models predictions using the \texttt{PC+PV} dataset on the Italian, German, Belgian and Dutch influenza incidence. The dashed lines delimits the various influenza seasons.}
\label{fig:model-prediction-pageviews-pagecounts}

\end{figure*}

Figures~\ref{fig:model-prediction-pageviews} and~\ref{fig:model-prediction-pageviews-pagecounts} show the predicted influenza trends for each of the given countries when using \texttt{PV} and \texttt{PC+PV}, respectively. Sometimes the models produce visible ``spikes''. Those outliers are related to a sudden page views increment (or decrement) of one or more of the feature pages.
These sudden changes in page views counts could be caused by many factors. For instance, unexpected news coverage of a certain topic may artificially increase the total views of a certain page, thus leading to overestimation of the results.

Ideally, we want our model to be robust against these kinds of situation and we prefer models which generate less ``spikes''. This objective reflects also on the set of pages we are using to monitor ILI incidence. We would like our model to select highly informative pages, which however do not present sudden changes in the page views count.
We can see from the graphs how the \categories model (green line) generates many of these unwanted values. On the contrary, \cyclerank and \ppagerank are more resilient and produces better predictions. Again, by using the \texttt{PC+PV} dataset as training set we can increase the robustness of all the models thus making them less susceptible to sudden page views changes.

\begin{table}[!t]
\centering
\caption{Mean Pearson Correlation Coefficient. The bold values indicate the best results over all combinations of models and datasets. The green color indicates the best results by either using the \texttt{PV} or the \texttt{PC+PV} dataset, by keeping fixed the country and the model (no color means no changes).}

\begin{tabular}{|c|cc|cc|cc|}
\hline
 & \multicolumn{2}{c|}{CycleRank} & \multicolumn{2}{c|}{PPageRank} & \multicolumn{2}{c|}{Categories} \\ \cline{2-7} 
\multirow{-2}{*}{Country} & \multicolumn{1}{c|}{PV} & PC+PV & \multicolumn{1}{c|}{PV} & PC+PV & \multicolumn{1}{c|}{PV} & PC+PV \\ \hline
IT & \cellcolor[HTML]{FFCCC9}0.898 & \cellcolor[HTML]{9AFF99}0.906 & \cellcolor[HTML]{FFCCC9}0.879 & \cellcolor[HTML]{9AFF99}\textbf{0.923} & \cellcolor[HTML]{FFCCC9}0.790 & \cellcolor[HTML]{9AFF99}0.913 \\
DE & \cellcolor[HTML]{FFCCC9}0.755 & \cellcolor[HTML]{9AFF99}0.865 & \cellcolor[HTML]{9AFF99}\textbf{0.914} & \cellcolor[HTML]{FFCCC9}0.874 & \cellcolor[HTML]{FFCCC9}0.482 & \cellcolor[HTML]{9AFF99}0.548 \\
NL & \cellcolor[HTML]{9AFF99}\textbf{0.757} & \cellcolor[HTML]{FFCCC9}0.681 & \cellcolor[HTML]{9AFF99}0.746 & \cellcolor[HTML]{FFCCC9}0.703 & \cellcolor[HTML]{FFCCC9}0.423 & \cellcolor[HTML]{9AFF99}0.657 \\
BE & \cellcolor[HTML]{FFCCC9}0.487 & \cellcolor[HTML]{9AFF99}0.793 & \cellcolor[HTML]{FFCCC9}0.625 & \cellcolor[HTML]{9AFF99}\textbf{0.809} & \cellcolor[HTML]{9AFF99}0.804 & \cellcolor[HTML]{FFCCC9}0.733 \\ \hline
\end{tabular}


\label{tab:experiments-results}
\end{table}

\begin{table}[!t]
\centering
\caption{Number of influenza peaks predicted correctly ($\#peaks = 4$). The bold values indicate the best results over all combinations of models and datasets. The first value in each cell indicates how many peaks were predicted correctly, the second value indicates how many peeks where predicted in the $\pm2$ weeks range. The green color indicates the best results by either using the \texttt{PV} or the \texttt{PC+PV} dataset, by keeping fixed the country and the model (no color means no changes).}
\begin{tabular}{|c|cc|cc|cc|}
\hline
 & \multicolumn{2}{c|}{CycleRank} & \multicolumn{2}{c|}{PPageRank} & \multicolumn{2}{c|}{Categories} \\ \cline{2-7} 
\multirow{-2}{*}{Country} & \multicolumn{1}{c|}{PV} & PC+PV & \multicolumn{1}{c|}{PV} & PC+PV & \multicolumn{1}{c|}{PV} & PC+PV \\ \hline
IT & 0 (4) & 0 (4) & \cellcolor[HTML]{FFCCC9}1 (3) & \cellcolor[HTML]{9AFF99}\textbf{1 (4)} & \cellcolor[HTML]{FFCCC9}0 (3) & \cellcolor[HTML]{9AFF99}\textbf{1 (4)} \\
DE & \cellcolor[HTML]{FFCCC9}0 (3) & \cellcolor[HTML]{9AFF99}\textbf{2 (4)} & \cellcolor[HTML]{FFCCC9}1 (4) & \cellcolor[HTML]{9AFF99}\textbf{2 (4)} & 1 (3) & 1 (3) \\
NL & \cellcolor[HTML]{FFFFFF}0 (3) & \cellcolor[HTML]{FFFFFF}0 (3) & \cellcolor[HTML]{9AFF99}0 (3) & \cellcolor[HTML]{FFCCC9}0 (2) & \cellcolor[HTML]{FFCCC9}0 (2) & \cellcolor[HTML]{9AFF99}0 (3) \\
BE & \cellcolor[HTML]{FFCCC9}0 (2) & \cellcolor[HTML]{9AFF99}0 (3) & \cellcolor[HTML]{FFCCC9}0 (2) & \cellcolor[HTML]{9AFF99}0 (3) & \cellcolor[HTML]{FFCCC9}0 (4) & \cellcolor[HTML]{9AFF99}\textbf{1 (3)} \\ \hline
\end{tabular}

\label{tab:experiments-results-peaks}

\end{table}

\subsection{Peak Accuracy}

The peak prediction performance can be seen in Table~\ref{tab:experiments-results-peaks}. The models are able to estimate the highest influenza peak within a 2-weeks range with respect to the ground-truth value. Unfortunately, they do not reach a level of accuracy high enough to estimate correctly the peak. For instance, on Dutch and Belgian data, the models do not estimate correctly any peak week, but they still position it in the two-weeks range. In this case, by examining the models trained with the extended dataset (\texttt{PC+PV}), in 8 models over 12 we can see a small increment in the correctly predicted peaks.  

\section{Feature Analysis}\label{sec:results-features}

In this section we analyze which features are selected by each model with the aim of characterizing how the chosen features affect the final estimate. Namely, which are the best predictors for estimating ILI incidence with Wikipedia? Are the selected predictors (Wikipedia's pages) related in some way to the \wiki{Influenza} topic? Is there a difference between using hand-picked features or automatically discovered ones? 

\begin{table*}[]
    \centering
    \caption{Analysis of the number of features selected by the models for each influenza season. We record the minimum, maximum and mean number of features for all the countries and for the different dataset used.}
    \resizebox{\textwidth}{!}{%
\begin{tabular}{|c|ccc|ccc|ccc|ccc|ccc|ccc|}
\hline
\multicolumn{1}{|l|}{\multirow{3}{*}{\textbf{Country}}} & \multicolumn{6}{c|}{\textbf{CycleRank}} & \multicolumn{6}{c|}{\textbf{PPageRank}} & \multicolumn{6}{c|}{\textbf{Categories}} \\ \cline{2-19} 
\multicolumn{1}{|l|}{} & \multicolumn{3}{c|}{PV} & \multicolumn{3}{c|}{PC+PV} & \multicolumn{3}{c|}{PV} & \multicolumn{3}{c|}{PC+PV} & \multicolumn{3}{c|}{PV} & \multicolumn{3}{c|}{PC+PV} \\ \cline{2-19} 
\multicolumn{1}{|l|}{} & \textit{Min} & \textit{Max} & \textit{Mean} & \multicolumn{1}{c}{\textit{Min}} & \multicolumn{1}{c}{\textit{Max}} & \multicolumn{1}{c|}{\textit{Mean}} & \textit{Min} & \textit{Max} & \textit{Mean} & \textit{Min} & \textit{Max} & \textit{Mean} & \textit{Min} & \textit{Max} & \textit{Mean} & \textit{Min} & \textit{Max} & \textit{Mean} \\ \hline
IT & 30 & 65 & 48.75 & 37 & 135 & 64.25 & 29 & 71 & 49.25 & 44 & 165 & 82.25 & 57 & 81 & 74.00 & 17 & 93 & 58.25 \\
DE & 27 & 71 & 55.25 & 84 & 96 & 89.25 & 22 & 51 & 37.50 & 69 & 123 & 101.00 & 34 & 75 & 61.00 & 75 & 89 & 81.75 \\
NL & 7 & 30 & 17.75 & 30 & 40 & 34.75 & 17 & 67 & 34.25 & 19 & 82 & 52.00 & 20 & 67 & 51.75 & 47 & 82 & 66.00 \\ 
BE & 8 & 39 & 23.75 & 29 & 34 & 31.75 & 20 & 67 & 42.50 & 37 & 56 & 48.50 & 16 & 70 & 38.50 & 7 & 12 & 9.25 \\
\hline
\end{tabular}%
}

    \label{tab:min-max-mean-features}
\end{table*}

\subsection{Top-5 Wikipedia's Predictors}

First, we focus on the estimated relative importance of each predictor. We wanted to find which pages are the most useful to estimate the ILI incidence of a given week. We focused on the weights assigned to the various Wikipedia's pages by the models. We directed our attention to those features which received a positive weight. A positive weight means that there is a positive correlation between the predictor and the influenza incidence. 
To detect the most successful subset of features for the final model, we ranked the features based on their mean weight over the different models. If a given Wikipedia's page is selected multiple times by separate models for different influenza seasons, it means that we found a feature which can be used to predict the ILI incidence over many seasons.  

We collected the top-5 features found by each model type (\cyclerank, \ppagerank and \textit{Categories}) and for each dataset, \texttt{PV} and \texttt{PC+PV}. We also computed the shortest path distance $D_I$ between the \wiki{Influenza} page to the found predictor and the Pearson Correlation Coefficient (PCC) for each chosen page against the influenza incidence. The complete results can be seen in Table \ref{tab:top-5-wikipedia-pages}. 
From the results, we can notice how the top-5 predictors of many models are Wikipedia's pages related to the influenza topic. If we perform a rapid semantic analysis, some of them refer directly to the symptoms (e.g., \wiki{Fever}, \wiki{Bronchitis}, \wiki{Chills}), the pathogens (e.g., \wiki{Influenza A virus subtype H1N1}) or to similar concepts related to ILI (e.g., \wiki{Peramivir}, \wiki{Swine Influenza}).
We can also identify some super-predictors. Namely, some features which are chosen multiple times by several models (e.g., \wiki{Febbre} for the Italian models or \wiki{Grippe} for both the Belgian and Dutch models). These super-predictors tend to have a very high PCC. Thus they are very valid in predicting the variation of the ILI incidence. Ultimately, 9 out the best 12 models considered in Table \ref{tab:top-5-wikipedia-pages} show to have at least half of their top-5 predictors with a distance $D_I$ less or equal than 2 from the \wiki{Influenza} page.  More interestingly, the \textit{Categories} models show the same behaviour, thus hinting that there may be some relationship between the optimal predictors and their position in Wikipedia's graph.

\subsection{Size of the feature subsets}

Table~\ref{tab:min-max-mean-features} shows some statistics about how many features were selected by each model during training. We record the minimum, maximum and mean value of the total features chosen by all the models. Generally, we can see how the final models always produce subsets of the original feature sets.
Not all the Wikipedia's pages available are good predictors and selecting too many features would cause loss of generalization. 
Models trained with the \texttt{PC+PV} dataset show a higher mean than their \texttt{PV} counterparts. This can be explained by the fact that by employing a larger dataset we may need more predictors to accurately fit the training set. Ultimately, we can see how the \cyclerank models tend to choose a smaller feature subset for all the countries still retaining the same accuracy.
In conclusion, we argue that we need to monitor just a few pages to be able to obtain good estimates of the ILI incidence.  

\subsection{Features shared between models}

\begin{figure*}
\centering

\subfloat{
\includegraphics[width=0.5\linewidth]{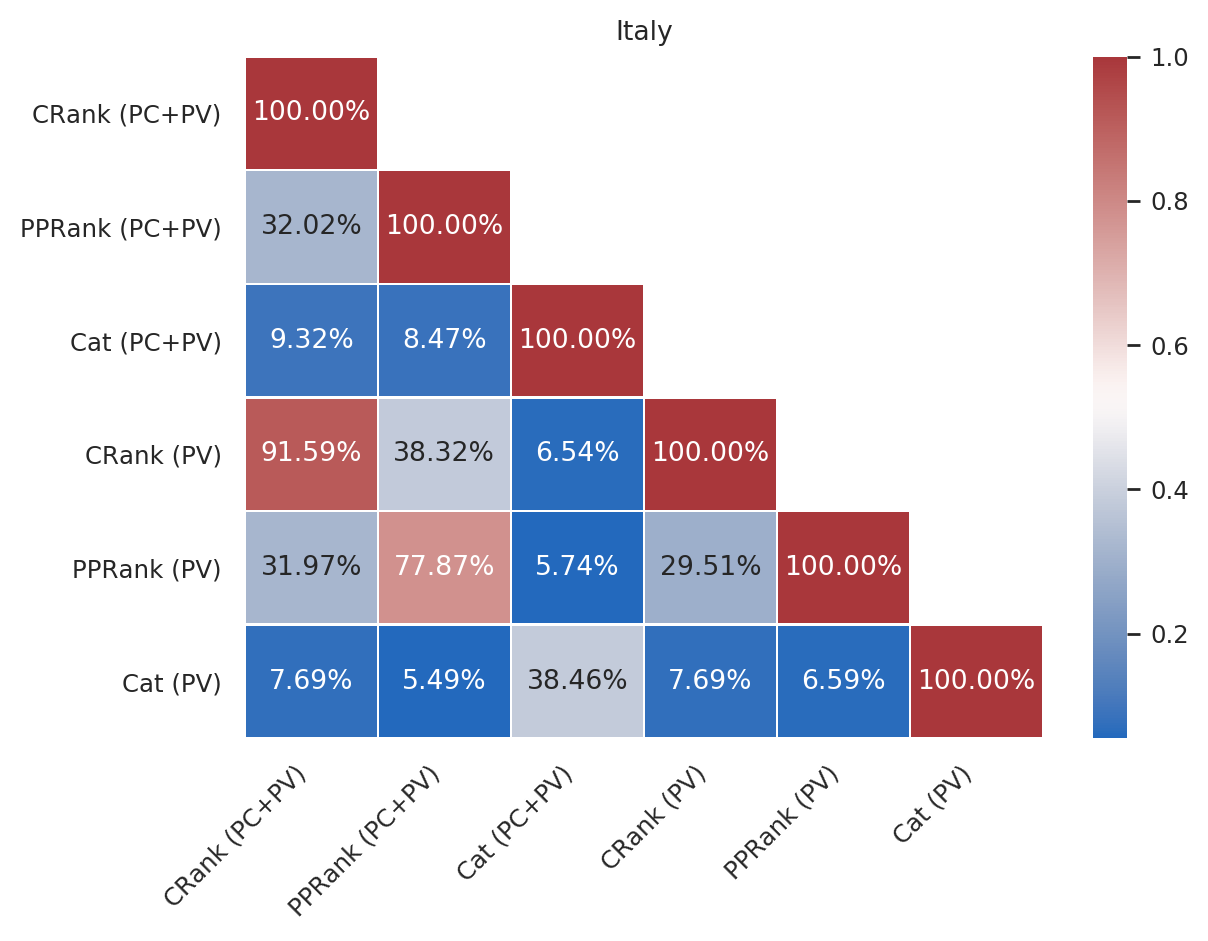}
}%
\subfloat{
\includegraphics[width=0.5\linewidth]{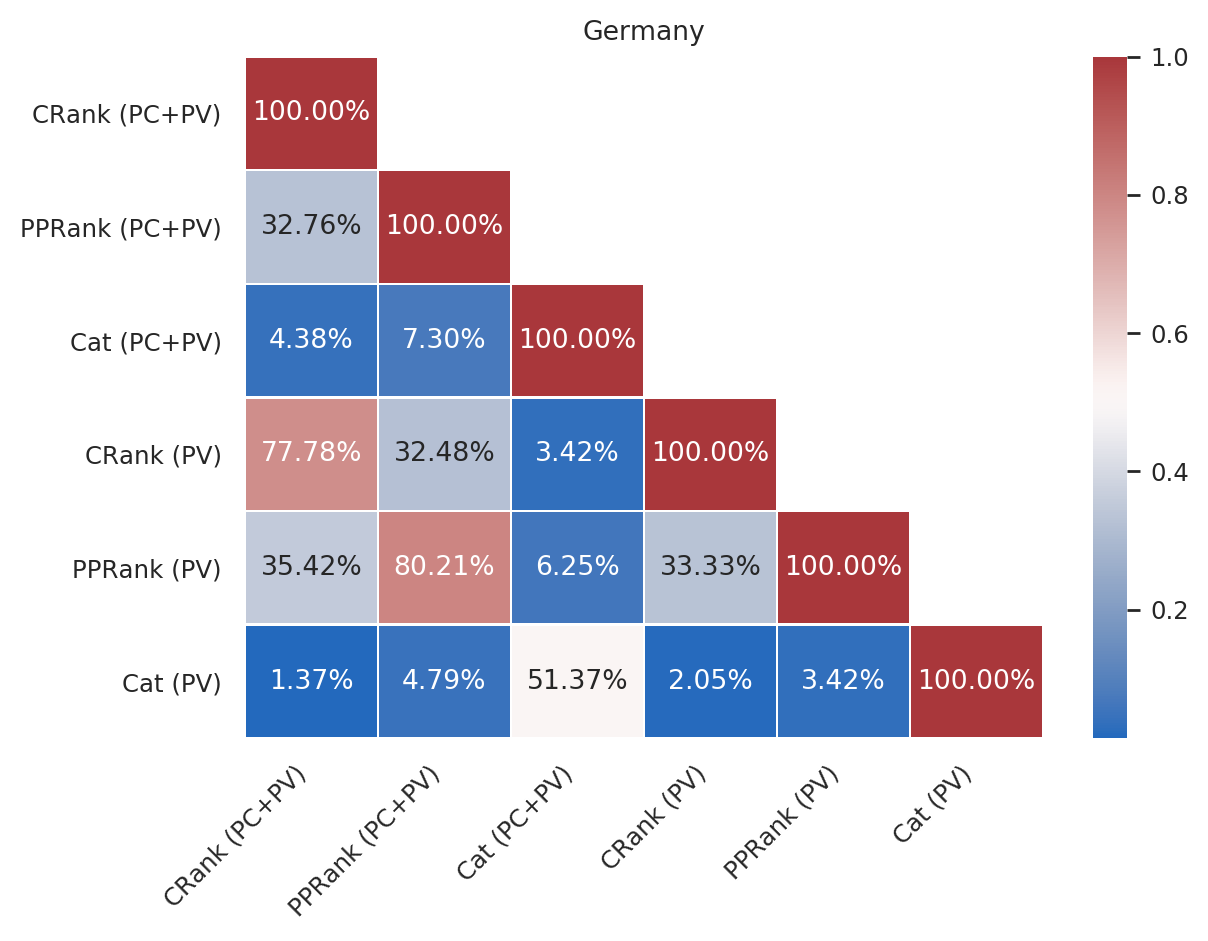}
}

\subfloat{
\includegraphics[width=0.5\linewidth]{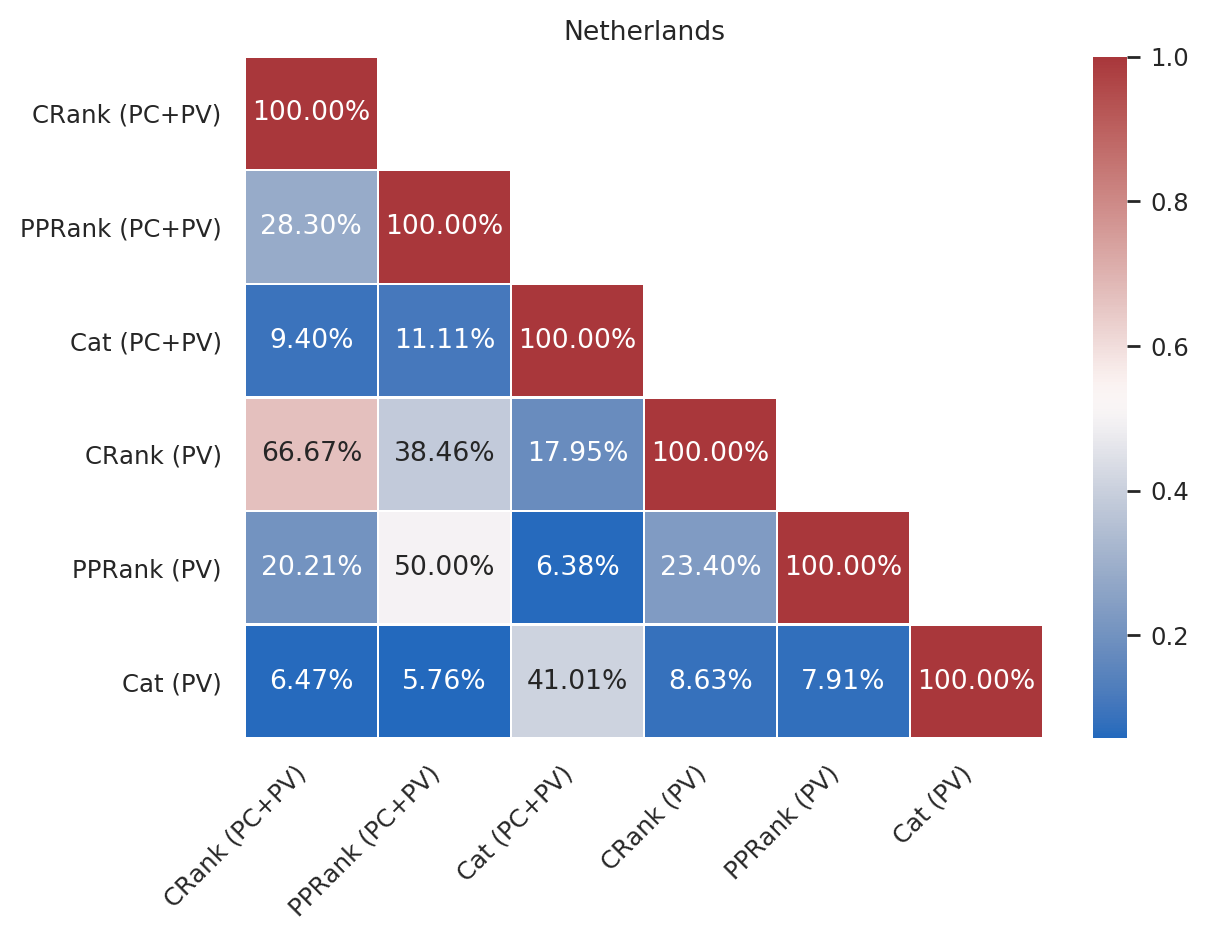}
}%
\subfloat{
\includegraphics[width=0.5\linewidth]{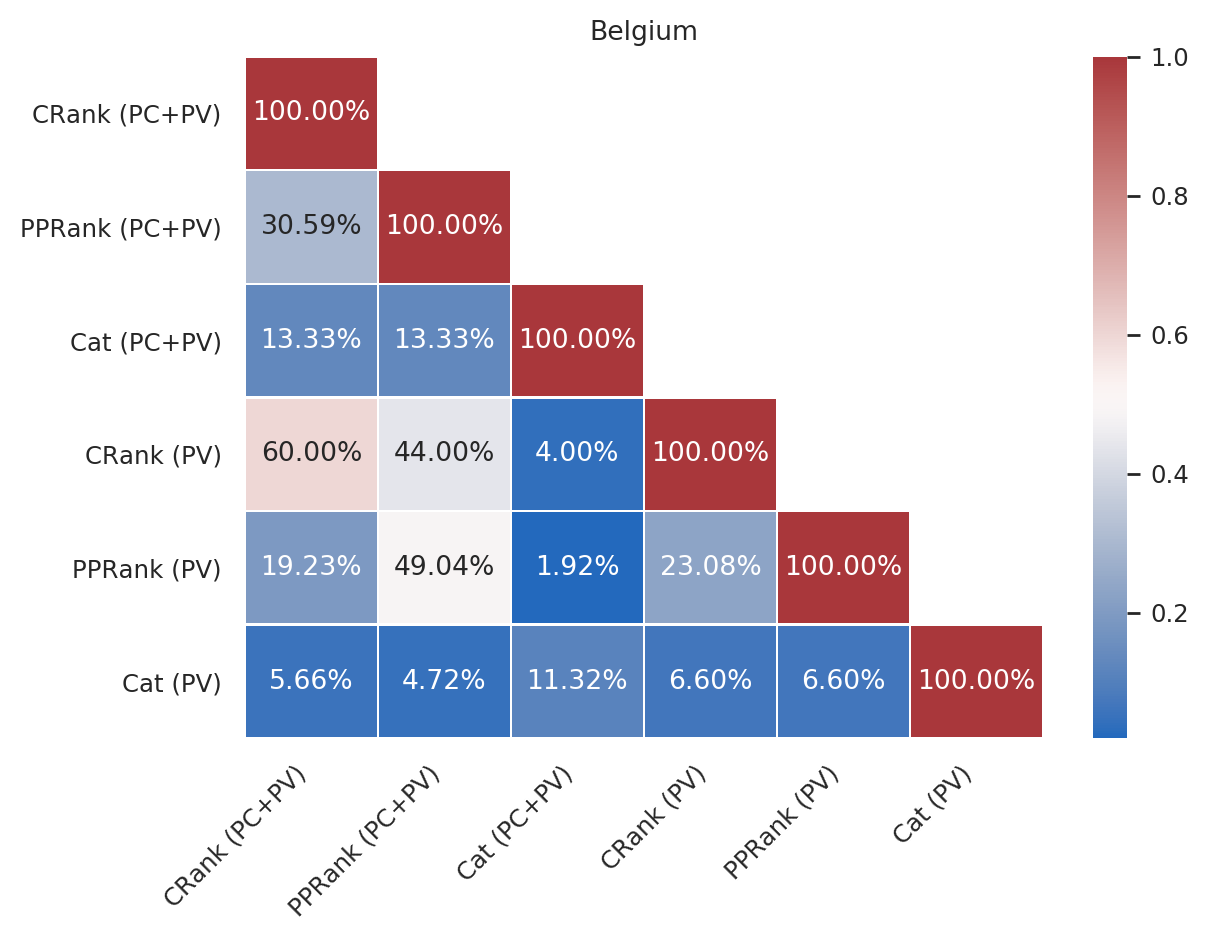}
}

\caption{Percentage of features shared by each model. We compare how many features are shared by each model (CRank=\cyclerank, PPRank=\ppagerank, Cat=\categories). We also report the percentage of feature shared by models trained with the two different dataset, \texttt{PV} and \texttt{PC+PV}. We recorded all the features selected by each model for each influenza season (a feature was included if and only its weight in at least one model was different than zero). Then, we computed the intersection between each set of features for each combination model/dataset.}
\label{fig:common-features-models}

\end{figure*}

We were also interested in investigating internal predictor variations which may occur when training models by changing the initial feature set. Figure \ref{fig:common-features-models} shows how many features are shared by the final trained models. 
First of all, if we compare the models trained with both the available datasets, \texttt{PV} and \texttt{PC+PV}, but with fixed feature set (e.g., using only \cyclerank), we can notice how they will both select identical features. The \cyclerank models seem to be less affected (more than $60\%$ shared features between \texttt{PV} and \texttt{PC+PV}), while \ppagerank and \textit{Categories} show a higher degree of variation. 
This fact is interesting because it tells us that providing a better dataset with more influenza seasons, such as \texttt{PC+PV}, affects the final predictors and it can push the models toward choosing different Wikipedia's pages. These changes, in turn, could lead to an improvement in the performances. 

In all the other cases, we can notice how the various models overlap just in some features. The models end up selecting almost non-intersecting sets of Wikipedia's voices. For instance, \categories shares only up to $13\%$ of the final predictors with \cyclerank and \ppagerank, regardless of the training dataset used.  Despite this, we can retain comparable performances. Again, this behaviour suggests that there are just a few Wikipedia's pages worth to monitor to estimate the ILI incidence.

\begin{table*}[th!]
    \caption{Top-5 Wikipedia's pages selected by the models for the influenza seasons from 2015 to 2019 for all the examined countries. Since each model could have been trained with either the PV or PV+PC dataset, we report only the models which performed better (see Table \ref{tab:experiments-results} for the best models). For each model, we report the page name, the shortest-path distance $D_I$ between the page and the corresponding \wiki{Influenza} page and the Pearson Correlation Coefficient (PCC) measured against the influenza incidence. We also report the corresponding page in the Engligh Wikipedia between parentheses. The value NE is used to specify when a page has no english equivalent. The value $D_I > 3$ indicates that the page is more than three hops away from the \wiki{Influenza} page.}
    \centering
    \subfloat[Italy]{
\resizebox{\textwidth}{!}{%
\begin{tabular}{|p{5cm}|c|c|p{5cm}|c|c|p{5cm}|c|c|}
\hline
\multicolumn{3}{|c|}{\textbf{Cyclerank}} & \multicolumn{3}{c|}{\textbf{PPageRank}} & \multicolumn{3}{c|}{\textbf{Categories}} \\ \hline
\multicolumn{1}{|c|}{Page Name} & \multicolumn{1}{c|}{\textit{PCC}} & \multicolumn{1}{l|}{$D_I$} & \multicolumn{1}{c|}{Page Name} & \multicolumn{1}{c|}{\textit{PCC}} & \multicolumn{1}{l|}{$D_I$} & \multicolumn{1}{c|}{Page Name} & \multicolumn{1}{c|}{\textit{PCC}} & \multicolumn{1}{c|}{$D_I$} \\
\hline
Febbre (Fever) & 0.55 &    1 &                                                    Febbre (Fever) & 0.55 &    1 &                                                    Febbre (Fever) & 0.55 &    1 \\
                        Oseltamivir (Oseltamivir) & 0.36 &    1 &  Influenzavirus A sottotipo H1N1 (Influenza A virus subtype H1N1) & 0.76 &  > 3 &  Influenzavirus A sottotipo H1N1 (Influenza A virus subtype H1N1) & 0.76 &  > 3 \\
                           Bronchite (Bronchitis) & 0.39 &    1 &                                         Oseltamivir (Oseltamivir) & 0.36 &    1 &                                            Bronchite (Bronchitis) & 0.39 &    1 \\
             Tosse post-virale (Post-viral cough) & 0.78 &  > 3 &                                            Bronchite (Bronchitis) & 0.39 &    1 &     Influenza asiatica (Influenza A virus subtype H2N2\#Asian flu) & 0.67 &  > 3 \\
 Vaccino influenza stagionale (Influenza vaccine) & 0.03 &  > 3 &                                             Polmonite (Pneumonia) & 0.45 &    1 &                                        Fago lambda (Lambda phage) & 0.56 &  > 3 \\
\hline
\end{tabular}%
}}

\subfloat[Germany]{
\resizebox{\textwidth}{!}{%
\begin{tabular}{|p{5cm}|c|c|p{5cm}|c|c|p{5cm}|c|c|}
\hline
\multicolumn{3}{|c|}{\textbf{Cyclerank}} & \multicolumn{3}{c|}{\textbf{PPageRank}} & \multicolumn{3}{c|}{\textbf{Categories}} \\ \hline
\multicolumn{1}{|c|}{Page Name} & \multicolumn{1}{c|}{\textit{PCC}} & \multicolumn{1}{l|}{$D_I$} & \multicolumn{1}{c|}{Page Name} & \multicolumn{1}{c|}{\textit{PCC}} & \multicolumn{1}{l|}{$D_I$} & \multicolumn{1}{c|}{Page Name} & \multicolumn{1}{c|}{\textit{PCC}} & \multicolumn{1}{c|}{$D_I$} \\
\hline
                                        Influenza-Schnelltest (NE) & 0.94 &  1 &                        Influenza-Schnelltest (NE) &  0.94 &    1 &    Erkältung (Common cold) & 0.59 &    1 \\
   Rifttalfieber (Rift Valley fever) & 0.61 &  2 &                          Pathogen (Pathogenicity) & -0.13 &    2 &        Impfstoff (Vaccine) & 0.09 &    2 \\
 Schweineinfluenza (Swine influenza) & 0.60 &  2 &                   Totenschein (Death certificate) &  0.24 &  > 3 &  Bradykardie (Bradycardia) & 0.66 &    2 \\
               Peramivir (Peramivir) & 0.25 &  2 &  Asiatische Grippe (1957–1958 influenza pandemic) &  0.79 &  > 3 &     Schüttelfrost (Chills) & 0.70 &    1 \\
               Impfung (Vaccination) & 0.17 &  2 &                Pferdeinfluenza (Equine influenza) &  0.31 &    2 &        Rhinosinusitis (NE) & 0.23 &  > 3 \\
\bottomrule
\end{tabular}%
}}

\subfloat[Belgium]{
\resizebox{\textwidth}{!}{%
\begin{tabular}{|p{5cm}|c|c|p{5cm}|c|c|p{5cm}|c|c|}
\hline
\multicolumn{3}{|c|}{\textbf{Cyclerank}} & \multicolumn{3}{c|}{\textbf{PPageRank}} & \multicolumn{3}{c|}{\textbf{Categories}} \\ \hline
\multicolumn{1}{|c|}{Page Name} & \multicolumn{1}{c|}{\textit{PCC}} & \multicolumn{1}{l|}{$D_I$} & \multicolumn{1}{c|}{Page Name} & \multicolumn{1}{c|}{\textit{PCC}} & \multicolumn{1}{l|}{$D_I$} & \multicolumn{1}{c|}{Page Name} & \multicolumn{1}{c|}{\textit{PCC}} & \multicolumn{1}{c|}{$D_I$} \\
\hline
                                                            Koorts (Fever) & 0.45 &    1 &                    Koorts (Fever) & 0.45 &    1 &               Griep (Influenza) & 0.59 &    1 \\
                          Griep (Influenza) & 0.59 &    1 &                 Griep (Influenza) & 0.59 &    1 &  Cytokinestorm (Cytokine storm) & 0.35 &    2 \\
 Mexicaanse griep (2009 swine flu pandemic) & 0.56 &  > 3 &           Chordadieren (Chordate) & 0.33 &  > 3 &       Flavivirus (Flaviviridae) & 0.33 &  > 3 \\
                  Azerbeidzjan (Azerbaijan) & 0.29 &    2 &                 1950-1959 (1950s) & 0.18 &  > 3 &                 Salkvaccin (NE) & 0.04 &  > 3 \\
                     Virulentie (Virulence) & 0.03 &    2 &  Antigene drift (Antigenic drift) & 0.33 &  > 3 &       Lassakoorts (Lassa fever) & 0.39 &  > 3 \\
\bottomrule
\end{tabular}%
}}

\subfloat[Netherlands]{
\resizebox{\textwidth}{!}{%
\begin{tabular}{|p{5cm}|c|c|p{5cm}|c|c|p{5cm}|c|c|}
\hline
\multicolumn{3}{|c|}{\textbf{Cyclerank}} & \multicolumn{3}{c|}{\textbf{PPageRank}} & \multicolumn{3}{c|}{\textbf{Categories}} \\ \hline
\multicolumn{1}{|c|}{Page Name} & \multicolumn{1}{c|}{\textit{PCC}} & \multicolumn{1}{l|}{$D_I$} & \multicolumn{1}{c|}{Page Name} & \multicolumn{1}{c|}{\textit{PCC}} & \multicolumn{1}{l|}{$D_I$} & \multicolumn{1}{c|}{Page Name} & \multicolumn{1}{c|}{\textit{PCC}} & \multicolumn{1}{c|}{$D_I$} \\
\hline
                          Griep (Influenza) &  0.75 &    1 &              Griep (Influenza) & 0.75 &    1 &                                           Griep (Influenza) & 0.75 &    1 \\
                   Hongkonggriep (Hong Kong flu) &  0.51 &    1 &  Hongkonggriep (Hong Kong flu) & 0.51 &    1 &                                              Koorts (Fever) & 0.57 &    1 \\
 Aziatische griep (1957–1958 influenza pandemic) &  0.61 &  > 3 &          Influenzavirus A (NE) & 0.38 &  > 3 &                  Mexicaanse griep (2009 swine flu pandemic) & 0.43 &  > 3 \\
                           Influenzavirus A (NE) &  0.38 &  > 3 &            Chirurgie (Surgery) & 0.25 &    2 &                              Cytokinestorm (Cytokine storm) & 0.48 &    2 \\
                         Amantadine (Amantadine) & -0.17 &    1 &                Immuniteit (NE) & 0.14 &  > 3 &  Respiratoir syncytieel virus (Respiratory syncytial virus) & 0.07 &  > 3 \\
\bottomrule
\end{tabular}%
}}
    \label{tab:top-5-wikipedia-pages}
\end{table*}

\section{Discussion}\label{sec:discussion}


The methods outlined in the previous sections complement more traditional techniques, especially if these conventional approaches present a lag between the collection of the data and the publications of the results. For instance, some illnesses require to conduct some laboratory tests such to have a positive/negative response. In such scenarios, it could be beneficial to employ these Wikipedia-based solutions to gain rapidly additional (albeit maybe less precise) information. 
More importantly, we have shown the feasibility of \textbf{using automatic methods to extract relevant predictors from Wikipedia itself by exploiting Wikipedia's links structure}. Wikipedia's structure is highly flexible, and its pages could be removed or changed very frequently, thus making some pages less relevant for the ILI estimation. Therefore, we may need to periodically retrain the models by incorporating new influenza data and additional Wikipedia's pages to ensure proper stability. We have seen how just a small subset of pages is responsible for a correct ILI estimation. Therefore we need to be able to find them correctly. \cyclerank and \ppagerank allow us to select new subsets of pages automatically, thus reducing the chances to have a low-performing model. Besides, we can employ these methods to effortlessly deploy machine learning models in other languages, thus making it easy to predict ILI incidence in different countries, as long as there is a Wikipedia's version with the same target language. Previously, this would have required to gather a team of expert in the target language in order to create a new feature dataset. Ultimately, we argue that this procedure is suitable to be applied to other illnesses, but further studies and analysis are required. 

We discuss here some of the main shortcomings of our proposed approach that we leave in future work. Firstly, we employed simple linear regression, since it is easier to train, and it provides interpretable results. However, by using linear models, we are implying that we can obtain the response variable, the ILI incidence, from the linear combination of the predictors. This assumption does not hold very well in practice, and more complex regression models are needed, such as Poisson regression or Negative Binomial regression.
Secondly, one of the main issues of Wikipedia-based models is that they are sensitive to the media coverage of the target topics. This excessive coverage could lead to overestimation of flu incidence because of the rising public attention. In our opinion, this is one of the primary concern and limit of this class of methods which relies on human-generated web traffic. This issue was also the reason behind the failure of several other similar solutions, like Google Flu Trends~\cite{lazer2014parable, santillana2014can}. 
We can account for these concerns by extending our model to take into account the media coverage for the \wiki{Influenza} topic.  We could perform anomaly detection on the page views, and we could instruct the model to normalize the various features to account for increased traffic. However, it is not trivial to monitor media and the news. We would need to identify reliable sources, and they should provide open dataset as Wikipedia does. Moreover, we may need to employ natural language processing analysis to check if the news subject is semantically related to ILI. This will require further tests and investigation.

\section{Conclusion}\label{sec:conclusions}

The recent pandemics have shown the importance of being able to estimate the spread of diseases in a fast and reliable way. In our case, we focused on the estimation of Influenza-Like Illnesses in European countries by using machine learning models, and the page views coming from Wikipedia, the online encyclopedia. We extended upon the state of the art by tackling the problem of using different language editions of Wikipedia that are visited by users in different European countries. 
Wikipedia is the first stop for many of health-related searches, when people are sick and look for information on the internet, they are likely to end up reading Wikipedia.
In this work, we have shown that we can exploit this information to build linear models to do nowcasting of the incidence of ILI in a given country. The incidence estimation concerns both the total number of cases and influenza peak detection, namely, the week in which we will see the highest number of infected people in all the influenza season. We can use these results to direct possible efforts of the public authorities and to devise safety measures in concert with more traditional approaches (e.g., SIR models, collection of laboratory tests, etc.). We tested our method on the previous four influenza seasons (from 2015 to 2019). More importantly, we have also shown how it is possible to use two automated methods, \cyclerank and \ppagerank, to extract automatically and rapidly highly-relevant features from Wikipedia by looking at Wikipedia's links structure. In previous works, this time-consuming task relied upon a group of experts physicians who manually chose every single relevant Wikipedia's page. The main advantage of \cyclerank and \ppagerank methods relies on their being able to generalize over multiple languages (and countries) without the need for expert supervision. We compared the performance of these novel methods against a more traditional one, called \textit{Categories}, which uses hand-picked pages. We showed that the models trained with these newly found features are equal or even outperform the models trained with hand-picked Wikipedia's pages.

\section*{Acknowledgments}
The authors would like to thank Michele Bortolotti and the team of ``Gestione Sistemi'' at the University of Trento for their support with the HPC cluster.

\bibliographystyle{ACM-Reference-Format}

\end{document}